\documentclass[aps,reprint,superscriptaddress,prl]{revtex4-1}
\usepackage{graphicx}
\usepackage{amsmath}
\usepackage{amssymb}
\usepackage[usenames,dvipsnames]{xcolor}
\usepackage{lipsum}
\usepackage{float}
\newcommand{\minus}{\scalebox{0.75}[1.0]{$-$}}

\usepackage{changepage}
\usepackage{tikz}
\usetikzlibrary{arrows.meta}
\usepackage{contour}
\usepackage[left]{lineno}
\usepackage{blindtext}
\definecolor{ashgray}{rgb}{0.7,0.75,0.71}
\definecolor{mspringgreen}{rgb}{0, 0.8, 0.1}
\definecolor{auburn}{rgb}{0.43, 0.21, 0.1}
\definecolor{ao(english)}{rgb}{0.0, 0.5, 0.0}
\definecolor{afw}{rgb}{0.95, 0.95, 0.96}
\definecolor{magnolia}{rgb}{0.97, 0.96, 1.0}
\definecolor{wsmk}{rgb}{0.96, 0.96, 0.96}
\newcommand{\CommaPunct}{\mathpunct{\raisebox{0.5ex}{,}}}
\newcommand{\cca}{${\rm CaCo_{2-y}As_2}$}
\begin{document}
\title{Effective One-Dimensional Coupling in the Highly-Frustrated Square-Lattice Itinerant Magnet CaCo$_{\mathrm{2}-y}$As$_{2}$}
\author{A. Sapkota}
\author{B. G. Ueland}
\affiliation {Ames Laboratory, U.S. Department of Energy, Iowa State University, Ames, Iowa 50011, USA}
\affiliation{Department of Physics and Astronomy, Iowa State University, Ames, Iowa 50011, USA}
\author{V. K. Anand}
\affiliation {Ames Laboratory, U.S. Department of Energy, Iowa State University, Ames, Iowa 50011, USA}
\affiliation{Helmholtz-Zentrum Berlin f\"{u}r Materialien und Energie GmbH, Hahn-Meitner Platz 1, D-14109 Berlin, Germany}
\author{N. S. Sangeetha}
\affiliation {Ames Laboratory, U.S. Department of Energy, Iowa State University, Ames, Iowa 50011, USA}
\affiliation{Department of Physics and Astronomy, Iowa State University, Ames, Iowa 50011, USA}
\author{D. L. Abernathy}
\author{M. B. Stone}
\affiliation{Quantum Condensed Matter Division, Oak Ridge National Laboratory, Oak Ridge, Tennessee 37831, USA}
\author{J. L. Niedziela}
\affiliation{Instrument and Source Division, Oak Ridge National Laboratory, Oak Ridge, Tennessee 37831, USA}
\author{D. C. Johnston}
\author{A. Kreyssig}
\author{A. I . Goldman}
\author{R. J. McQueeney}
\affiliation {Ames Laboratory, U.S. Department of Energy, Iowa State University, Ames, Iowa 50011, USA}
\affiliation{Department of Physics and Astronomy, Iowa State University, Ames, Iowa 50011, USA}
%\author{A. Sapkota$^{1,2}$, B. G. Ueland$^{1,2}$, V. K. Anand$^{1,2,3}$, D. L. Abernathy$^{4}$, M. B. Stone$^{4}$, J. L. Niedziela$^{5}$, D. C. Johnston$^{1,2}$, A. Kreyssig$^{1,2}$, A. I. Goldman$^{1,2}$ and R. J. McQueeney$^{1,2}$}

%\linenumbers
%\maketitle
%\begin{affiliations}
%\item{Ames Laboratory, U.S. DOE, Iowa State University, Ames, Iowa 50011, USA}
%\item{Department of Physics and Astronomy, Iowa State University, Ames, Iowa 50011, USA}
%\item{Helmholtz-Zentrum Berlin f\"{u}r Materialien und Energie GmbH, Hahn-Meitner Platz 1, D-14109 Berlin, Germany}
%\item{Quantum Condensed Matter Division, Oak Ridge National Laboratory, Oak Ridge, Tennessee 37831, USA}
%\item{Instrument and Source Division, Oak Ridge National Laboratory, Oak Ridge, Tennessee 37831, USA}
%\end{affiliations}
%

\begin{abstract}
 Inelastic neutron scattering measurements on the itinerant antiferromagnet (AFM) \cca\ at a temperature of 8~K reveal two orthogonal planes of scattering perpendicular to the Co square lattice in reciprocal space, demonstrating the presence of effective one-dimensional spin interactions.  These results are shown to arise from near-perfect bond frustration within the $J_1$-$J_2$ Heisenberg model on a square lattice with ferromagnetic $J_1$, and hence indicate that the extensive previous experimental and theoretical study of the $J_1$-$J_2$ Heisenberg model on local-moment square spin lattices should be expanded to include itinerant spin systems.
\end{abstract}

\maketitle
Magnetic frustration arises when competing interactions between magnetic moments (spins) cannot be mutually satisfied.$~$It suppresses the development of long-range magnetic order, and often creates enhanced spin fluctuations, which can lead to a variety of novel phases including quantum spin liquids \textcolor{blue}{\cite{Leon_2010,Savary_2017}}, spin and electronic nematic phases and unconventional superconductivity \textcolor{blue}{\cite{Chandra_1990,Xu_2008,Fang_2008,Shanon_2006}}. There are many examples of materials in which geometry of the lattice leads to frustration, such as pyrochlore, spinel, or Kagom$\acute{\mathrm{\text{e}}}$ systems \textcolor{blue}{\cite{Leon_2010,Greedan_2001}}. However, in the case of a square lattice system, the frustration can arise from competing nearest-neighbor (NN) and next-nearest-neighbor (NNN) interactions \textcolor{blue}{\cite{Shannon_2004,Nath_2009}}.\par
 
Compounds with the chemical formula \textit{ATM}$_{2}$As$_{2}$, (with \textit{A} $=$ Ca, Sr, Ba and \textit{TM} $=$ Mn, Fe, Co, Ni), form a large class of quasi-two-dimensional (quasi-2D) materials containing layers of $TM$ ions on a square lattice, which are stacked along \textit{c}. Despite the crystal structure being three dimensional, they are considered quasi-2D for magnetism, as the interactions between layers are much smaller than those within the layers. Much of the recent motivation for the study of these materials is due to the proximity of antiferromagnetic (AFM) order and high-temperature superconductivity in the doped variants of \textit{TM} $=$ Fe compounds \textcolor{blue}{\cite{Sefat_2008,Jasper_2008,Kumar_2009,Johnston_2010}}. The \textit{ATM}$_{2}$As$_{2}$ materials adopt several different magnetic structures, including N$\acute{\mathrm{\text{e}}}$el-(or checkerboard)type  AFM (e.g.,$~$BaMn$_{2}$As$_{2}$ \textcolor{blue}{\cite{Johnston_2011}}), stripe-type AFM (e.g.,$~$\textit{A}Fe$_{2}$As$_{2}$ \textcolor{blue}{\cite{Johnston_2010}}), and A-type (e.g., \cca\ \textcolor{blue}{\cite{Quirinale_2013}}). The AFM order in the \textit{TM} $=$ Fe and Co variants is itinerant in nature, possessing an ordered moment of $\mu \lesssim1 \mu_{\textrm{B}}/$\textit{TM}.\par  
Despite their itinerant nature, the magnetic structures and spin fluctuations can be minimally described by considering NN ($J_{1}$) and NNN ($J_{2}$) magnetic exchange interactions between magnetic ions on a square lattice \textcolor{blue}{\cite{Jdef}}. In general, the magnetic ground state is determined by the relative strengths of $J_{1}$ and $J_{2}$, with N$\acute{\mathrm{\text{e}}}$el-type, stripe-type AFM, and FM/A-type  ordering occurring for $J_{1} > 2J_{2}$ ($J_{1} > 0$),  $|J_{1}| < 2J_{2}$ ($J_{2} > 0$),  and $\minus J_{1} > 2J_{2}$ ($J_{1} < 0$), respectively \textcolor{blue}{\cite{Shannon_2004,Johnston_2010}}. The system becomes frustrated and ordering in any of these magnetic structures is suppressed when $|J_{1}|\approx 2J_{2}$ and $J_{2} > 0$ (AFM). A frustration parameter, $\eta=J_{1}/2J_{2}$, quantifies the level of the magnetic frustration. Maximum frustration occurs when $\eta = 1$ ($ \minus 1$), and the stripe-type and N$\acute{\mathrm{\text{e}}}$el-type \big[stripe-type and ferromagnetic (FM)\big] ground states compete \textcolor{blue}{\cite{Shannon_2004,Johnston_2010}}.\par 

 The frustration parameter, $\eta$, can be determined experimentally by measuring the spatial anisotropy of the spin fluctuations using inelastic neutron scattering (INS). The spatial distributions of the spin fluctuations in different magnetic ground states depend strongly on $\eta$ as illustrated in the Supplemental Information (Fig.$~$\textcolor{blue}{S2}). For example, INS measurements for the parent and doped compositions of \textit{A}Fe$_{2}$As$_{2}$ compounds find $\eta$ in the range from 0.3-0.7 \textcolor{blue}{\cite{Tucker_2012,Harriger_2011}}, suggesting the presence of a moderate degree of frustration. INS experiments on SrCo$_{2}$As$_{2}$ observe stripe-type AFM spin fluctuations peaked at $\textbf{Q}_\mathrm{stripe}$ similar to \textit{A}Fe$_{2}$As$_{2}$ \textcolor{blue}{\cite{Jayasekara_2013}}. The anisotropy of the spin fluctuations gives $\eta \approx$ \minus0.6 \textcolor{blue}{\cite{Jayasekara_2013}}, and indicates a moderate frustration.\par  
 \setlength\belowcaptionskip{-2ex}
 \begin{figure*}[htb]
 	%\begin{tabular}{@{}l@{}l@{}}
 	\centering
 	\vspace{0em}{\includegraphics[trim = 25mm 82mm 25mm 91mm,clip,scale=0.9]{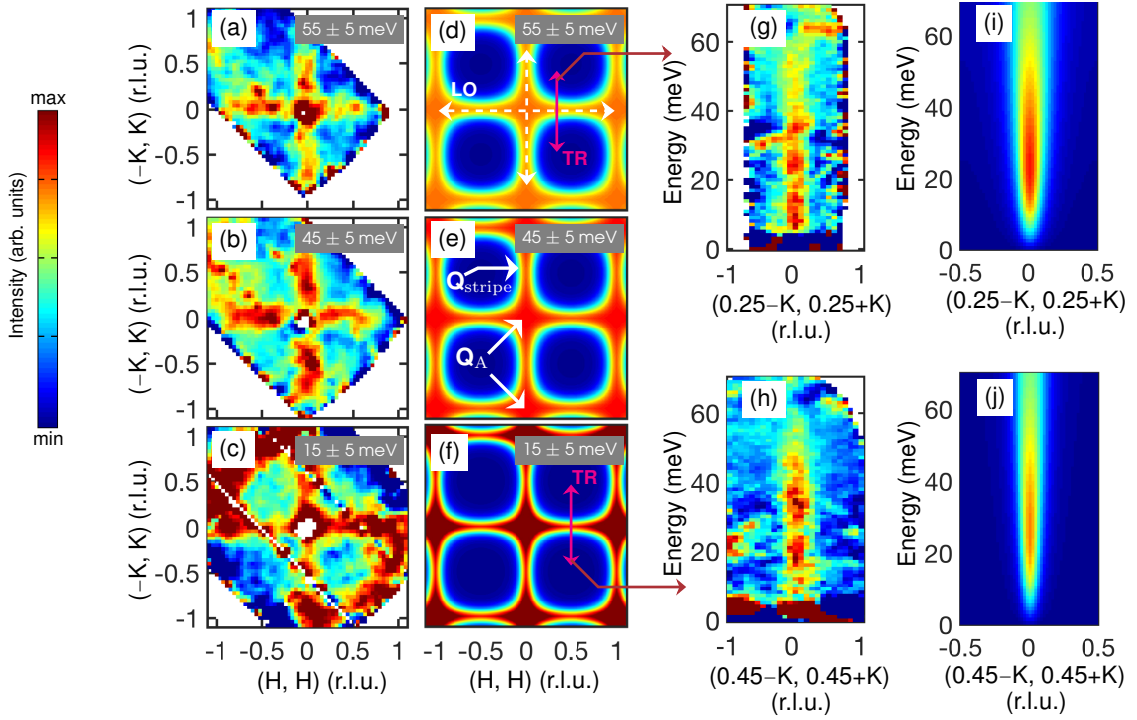}}
 	\caption{\noindent{Spin fluctuations in CaCo$_{\mathrm{2}-y}$As$_{2}$: Experimental INS Data (\textit{E$_{\mathrm{i}}$} $=$ 75 meV) vs Diffusive Model. (a)--(c), Constant-energy slices of the background subtracted data (the isotropic background is estimated similar to Ref.$~$\textcolor{blue}{\citenum{Tucker_2012}}) averaged over the energy ranges of 50--60 meV, 40--50 meV and 10--20 meV, respectively. (d)--(f) Corresponding constant-energy slices calculated using the diffusive model and parameters obtained from fitting the data shown in Fig.$~$\textcolor{blue}{2}. In Fig.$~$(d), the white dashed arrows show the LO directions and the maroon arrow shows the TR directions. (g),(h) Transverse slices of the background-subtracted data along the $[$\protect\minus\textit{K}, \textit{K}$]$ direction through (0.25, 0.25) and (0.45, 0.45) (r.l.u.), respectively, obtained after averaging over the LO direction by $\pm$ 0.1 r.l.u. (i),(j) Corresponding energy dependence of the scattering obtained using the diffusive model and parameters of fits to the data in Fig.$~$\textcolor{blue}{2}. The energy-dependent figures are averaged over all four symmetry-related quadrants. The color scale represents the intensity in each panel. The maximum/minimum intensities in the panels are (a) 0.34/\protect\minus0.5, (b) 0.42/\protect\minus0.3, (c) 0.65/\protect\minus0.15, (d)--(f) 0.43/0, (g) 0.52/\protect\minus0.45, (h) 0.65/\protect\minus0.35, (i) and (j) 0.52/0 in arbitrary units.}}
 \end{figure*}
Here, we report inelastic neutron scattering measurements of magnetic excitations in \cca\ at $T = 8$~K, below the AFM ordering temperature $T_{\rm N} = 52$~K \textcolor{blue}{\cite{Quirinale_2013,Anand_2014}}, that reveal {\it planes} of magnetic scattering in reciprocal space perpendicular to the Co square-lattice planes.\@ This is in sharp contrast to {\it rods} of scattering expected perpendicular to quasi-two-dimensional planes of spins but similar to the expectation for quasi-one-dimensional spin chains.  From this anisotropy, we model the data and estimate $\eta\approx -1$, corresponding to extreme frustration.\@ These observations indicate that previous extensive studies of the $J_1$-$J_2$ Heisenberg model on local-moment square spin lattices should be expanded beyond oxides to include itinerant spin systems found in arsenide and potentially other pnictide.  This has many potential interesting ramifications and applications.\par

 Single crystals of CaCo$_{\mathrm{2}-y}$As$_{2}$ ($y$ $\simeq$ 0.14), were grown using Sn flux \textcolor{blue}{\cite{Quirinale_2013,Anand_2014}}. The vacancies on the Co sites are randomly distributed, as x-ray and neutron diffraction measurements find no evidence for vacancy ordering \textcolor{blue}{\cite{Quirinale_2013,Anand_2014}}. INS experiments were performed on the ARCS Spectrometer \textcolor{blue}{\cite{Abernathy_2012}} at the Spallation Neutron Source, at Oak Ridge National Laboratory. Six single crystals of CaCo$_{\mathrm{2}-y}$As$_{2}$ with a total mass of $\mathtt{\sim}$1.1 g were co-aligned in the (\textit{H}, 0, \textit{L}) scattering plane (in tetragonal notation) with the full width at half maximum of less than 4$^{\circ}$. The co-aligned set was attached to the cold finger of a closed-cycle He cryostat. The measurements were carried out with the incident beam along the \textit{c} axis and incident energies of \textit{E$_{\mathrm{i}}$} $=$ 75 and 250 meV at \textit{T} = 8 K. Throughout the paper,  we define $ \textbf{Q} = (\textit{H\minus K, H+K, L}) = (2\pi/a)(H-K)\hat{\mathrm{\textbf{i}}} + (2\pi/a)(H+K)\hat{\mathrm{\textbf{j}}} + (2\pi/c)L\hat{\mathrm{\textbf{k}}}$ with respect to the tetragonal \textit{I}4/\textit{mmm} crystal system, where $ a = 3.98$~$\mathring{\mathrm{A}}$ and $c = 10.27$~$\mathring{\mathrm{A}}$. In this notation, the reciprocal lattice vectors for stripe and A-type magnetic order are $\textbf{Q}_\mathrm{stripe} = (\frac{1}{2},\frac{1}{2},1)+(m,n,l)$ with $m$, $n$ integers and $l$ even and $\textbf{Q}_\mathrm{A} = (H-K,H+K,L)$ where $H-K$ and $H+K$ are integers with $|H-K| +|H+ K|$ even and $L$ odd. The data were visualized using the Mslice software \textcolor{blue}{\cite{Mslice}}.\par

Figures \textcolor{blue}{1(a)--1(c), 1(g)} and \textcolor{blue}{1(h)} are the INS data measured on ARCS and show the presence of very striking spin fluctuations in CaCo$_{\mathrm{2}-y}$As$_{2}$. CaCo$_{\mathrm{2}-y}$As$_{2}$ has A-type AFM order at low temperatures and the low-energy spin fluctuations are expected to originate at $\textbf{Q}_\mathrm{A} =$ \big[(0, 0), (1, 1)\big], using 2D notation. However, constant energy slices of the data in Figs.$~$\textcolor{blue}{1(a)--(c)} show that the scattering from the spin fluctuations does not occur only around positions corresponding to $\textbf{Q}_\mathrm{A}$, but extends along the longitudinal (LO) direction, crossing positions including $\textbf{Q}_\mathrm{stripe} = $ (1/2, 1/2), at all accessible energies from $\sim$10--120 meV\@. In contrast, the scattering along the transverse (TR) direction is sharp. The energy dependence \big[Figs.$~$\textcolor{blue}{2(g), 2(h)} and Supplemental Material: Fig.$~$\textcolor{blue}{S3}\big] shows steep dispersion in the TR direction and the scattering extends in energy beyond 100 meV. As we will show below, the extreme anisotropy of the spin-fluctuations gives $\eta\approx\minus1$, implying that CaCo$_{\mathrm{2}-y}$As$_{2}$ exhibits near perfect magnetic frustration between FM/A-type and stripe-type ordering. \par

To describe the scattering data, we consider two models for A-type AFM. Using values of ${J}_{1}$ and ${J}_{2}$ appropriate for an A-type AFM ground state, we first use the linear spin-wave theory approximation to the Heisenberg model to calculate the neutron scattering cross-section corresponding to the values of \textbf{Q} and \textit{E} measured by INS. The details  and results of the model calculations are shown in the Supplemental Material: Fig.$~$\textcolor{blue}{S1}. We find that when adopting parameters corresponding to nearly maximal frustration ($\eta \approx \minus1$), the spin waves collapse along the LO direction, leading to a spin wave anisotropy that is similar to the experimental data at low energies. However, the calculated cross section and INS data show significant differences at higher energies (Supplemental Material: Fig.$~$\textcolor{blue}{S1}).\par

We next consider a model more appropriate for itinerant systems close to magnetic order. In itinerant magnets, the electronic degrees-of-freedom can result in a significant degree of Landau damping of the spin fluctuations due to the scattering of electrons. A diffusive model that describes such overdamped spin fluctuations in a nearly ordered system has been used to describe INS data on weakly ordered and metallic Ba(Fe$_{1-x}$Co$_{x}$)$_{2}$As$_{2}$ \textcolor{blue}{\cite{Inosov_2010,Tucker_2012,Tucker_2014}} and paramagnetic CaFe$_{2}$As$_{2}$ \textcolor{blue}{\cite{Diallo_2010}}. Neglecting the weak AFM interlayer interactions and keeping intralayer interactions up to NNN only, we develop a similar model for FM/A-type fluctuations. We find that the imaginary part of the generalized magnetic susceptibility, which is proportional to the INS spectrum, can be written as 

\begin{widetext}
\begin{equation}\label{a}
\chi'' (\textbf{q}+\textbf{Q$_\mathrm{A}$},\omega) = \dfrac{\chi_{\text{0}}\Gamma\omega}{\Gamma{^2}(1+\xi^{2}_{q}q^{2})^{2}+\omega^{2}}\CommaPunct
\end{equation}
where
\begin{equation}
\xi^{2}_{q}q^{2} = (4\xi/a)^{2}\Big\{\eta\big[\cos{(\dfrac{q_{x}+q_{y}}{2}a)}+\cos{(\dfrac{q_{y}\minus q_{x}}{2}a)}\big]+\cos{(\dfrac{q_{x}+q_{y}}{2}a)}\cos{(\dfrac{q_{y}\minus q_{x}}{2}a)}-2\eta-1\Big\}.
\end{equation}

\end{widetext}

 Here, $\chi_{\text{0}}$ is the static uniform susceptibility, $\xi$ is the magnetic correlation length, $\Gamma$ is the Landau damping parameter, and the $x$ and $y$ directions correspond to the [1 0 0] and [0 1 0] directions of the tetragonal \textit{I}4/\textit{mmm} crystal system, respectively.  In the itinerant picture relevant for the iron arsenides, $\eta$ arises from the spatial anisotropy of electronic velocities at the Fermi surface \textcolor{blue}{\cite{Tucker_2012,Knolle_2010}}.\par 
 \begin{figure}[htb]
 	\begin{adjustwidth}{0mm}{}
 		%\begin{tabular}{@{}l@{}l@{}}
 		\vspace{0em}
 		\centering	
 		\includegraphics[trim = 25mm 70mm 60mm 30mm,clip,width=9cm]{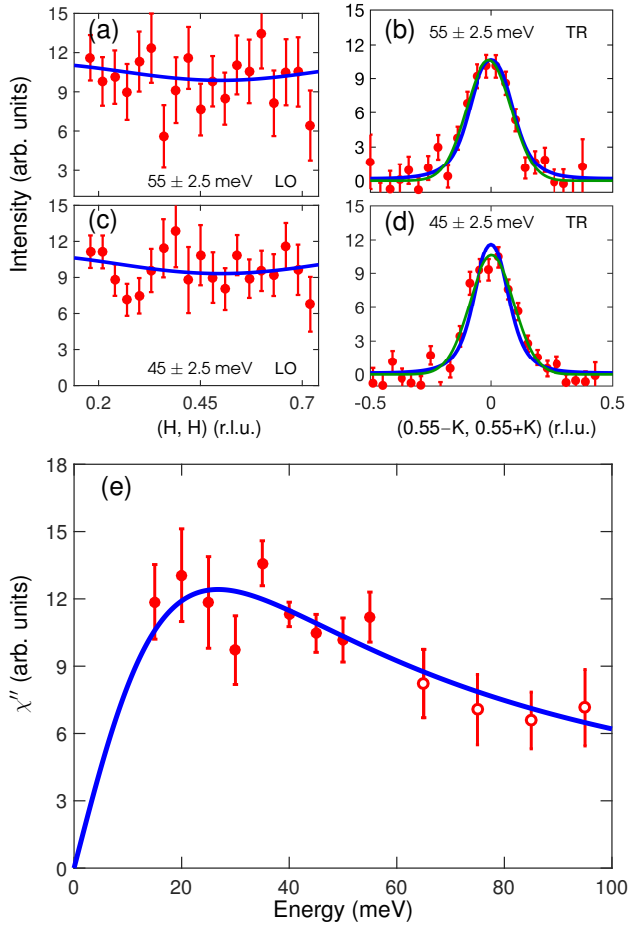}\vspace{0em}
 		%	\vspace{0em}\includegraphics[trim = 0mm 0mm 0mm 0mm,clip,width=20cm]{fig3_aftrrob}
 		%\renewcommand{\sfdefault}{phv}
 		%\renewcommand{\familydefault}{\sfdefault}\caption{\noindent\fontsize{11pt}{0pt}\selectfont{
 		\caption{\noindent{Constant-energy cuts and dynamical susceptibility of the data fitted with the diffusive model. (a),(c) LO and (b),(d) TR cuts of data through (0.55, 0.55) at (a), (b) 52.5--57.5 meV and (c), (d) 42.5--47.5 meV. TR cuts were averaged from 0.3--0.8 r.l.u.$~$in the LO direction and LO cuts $\pm$ 0.1 r.l.u. in the TR directions. Both data are corrected for the Co$^{+\text{1}}$ magnetic form factor after subtracting a background contribution estimated from Gaussian fits to the TR cuts. The green lines in the TR cuts are fits to a Gaussian lineshape and blue solid lines are fits to the diffusive model. (e) Energy dependence of the imaginary part of the dynamical susceptibility centered at (0.55, 0.55) and averaged along the LO direction from 0.3--0.8 r.l.u. Each data point is obtained after fitting the TR cuts with a Gaussian lineshape. The closed symbols are data measured with \textit {E$_{\mathrm{i}}$} $=$ 75 meV and open symbols with \textit{E$_{\mathrm{i}}$} $=$ 250 meV.}}
 	\end{adjustwidth}
 \end{figure}
 Similar to Ref.$~$[\textcolor{blue}{\citenum{Diallo_2010}}], we can experimentally determine $\eta=\dfrac{\xi_{\mathrm{LO}}^{2} + \xi_{\mathrm{TR}}^{2}}{\xi_{\mathrm{LO}}^{2} \minus \;\xi_{\mathrm{TR}}^{2}}$ in terms of the magnetic correlation lengths in the LO and TR directions. The values of the parameters in Eqs.$~$(\textcolor{blue}1) and (\textcolor{blue}2) were determined by fitting various cuts from the data as shown in Fig.$~$\textcolor{blue}{2}. The data are best fit with a constant ($\textbf{Q}$-independent) damping parameter \big[$\Gamma$~$=$~$20(4)$ meV\big]. The correlation lengths are much shorter in the LO direction (i.e. broader in reciprocal space) than in the TR direction, and we find that  $\eta = \minus1.03(2)$. This value of $\eta$ indicates the presence of extreme frustration in CaCo$_{\mathrm{2}-y}$As$_{2}$. The calculated neutron scattering cross sections, using the diffusive model, capture all of the essential features corresponding the INS data shown in Figs.$~$\textcolor{blue}{1} and \textcolor{blue}{2}.\par

 Also from the dispersion of A-type AFM in the TR direction \big[Fig.$~$\textcolor{blue}{2(a)}\big], we estimate that $S{J}_{1}$$\mathtt{\approx}$ \minus 57 meV, where $S$ is the magnitude of the spin \big[details of estimation are in Supplemental Material: Eq.$~$(\textcolor{blue}5)\big]. We find that the energy scale for the A-type AFM order is much smaller than the FM NN exchange energy ($k_{\text{B}}T_{\mathrm{N}}/S|J_{1}| \sim$ 0.1). Thus CaCo$_{\mathrm{2}-y}$As$_{2}$ orders at a temperature much lower than expected based on the strength of its magnetic interactions. This is a hallmark of strong magnetic frustration.\par
 
 \begin{figure}[htb!]
 	%\begin{adjustwidth}{0mm}{}
 	\vspace{0em}\hspace{0mm}
 	\centering
 	\includegraphics[trim = 62mm 150mm 53mm 24mm,clip,width=0.49\textwidth]{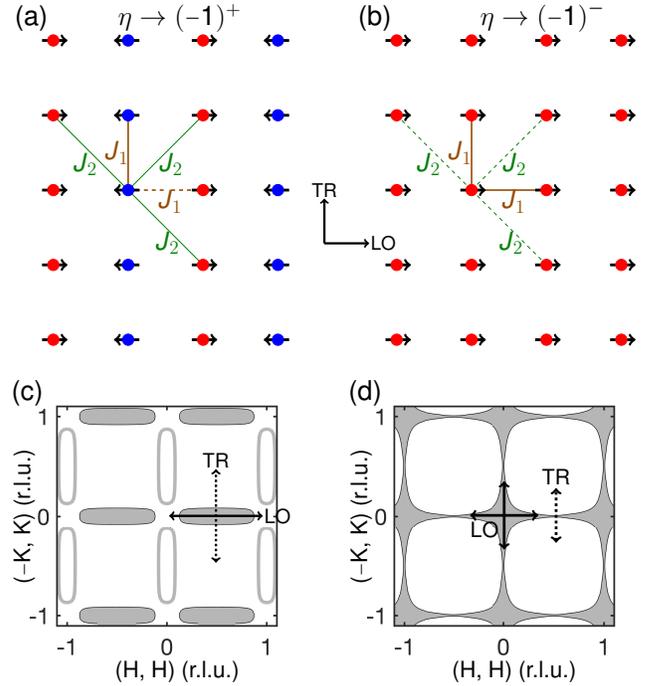}
 	%\put(-180,83){{\color{black}\fontsize{9}{9}\fontfamily{phv}\selectfont\colorbox{white}{\ovalbox{$\textbf{Q}_{\mathrm{\text{A}}}$}}}}
 	%\renewcommand{\sfdefault}{phv}
 	%\renewcommand{\familydefault}{\sfdefault}\caption{\noindent\fontsize{11pt}{0pt}\selectfont{
 	\caption{\noindent{(a),(b), Coupling of spins on a square lattice corresponding to stripe-type AFM and FM/A-type spin dynamics, respectively, for an extreme degree of frustration showing effective zero coupling along the LO directions. Red and blue circles represent transition metals on different magnetic sublattices. Magnetic NN (\textit{J}$_{1}$) and NNN (${\mathrm{\textit{J}}}_{2}$) interactions are shown where \textit{J}$_{1}$ is FM and \textit{J}$_{2}$ is AFM. Stable(frustrated) bonds are denoted by a solid (dashed) lines. (c),(d), Spin fluctuations corresponding to highly frustrated spin interactions in (a) and (b), respectively. The empty rods shown in (c) are due to the twin domain of (a), that has alternating chains of blue and red circles along the perpendicular directions. Both occur with equal populations.}}
 	%\end{adjustwidth}
 \end{figure}

Materials such as CaCo$_{\mathrm{2}-y}$As$_{2}$ are considered quasi-2D because interactions between Co layers are much smaller than those within the layer. In the case of CaCo$_{\mathrm{2}-y}$As$_{2}$, frustrated interactions within the layer reduce the dimensionality even further, leading to effectively 1D behavior. This is most easily pictured by considering \textit{J}$_{1}$ and ${\mathrm{\textit{J}}}_{2}$ interactions in a single square layer with the stripe-type AFM order, as shown in Fig.$~$\textcolor{blue}{3(a)}. The stripe-type magnetic structure consists of alternating chain of FM spins oriented in the TR direction \big[ordering vector ($1/2,1/2$)\big] with an effective interchain coupling of $\sim$ $|\textit{J}_{1}|$\minus2$\textit{J}_{2}$ which goes to zero as $\eta \rightarrow (\minus1)^+$. Moreover, the cost of flipping one FM chain against another goes to zero and the FM chains decouple. This results in a vanishing dispersion in the LO direction, but the steep dispersion remains in the TR direction as shown in Fig.$~$\textcolor{blue}{3(c)} where effective coupling is maximized as $\sim$ $\minus |\textit{J}_{1}|$\minus2$\textit{J}_{2}$. Rod-like fluctuations occur as a result of the effective zero coupling along the LO direction as shown in Fig.$~$\textcolor{blue}{3(c)}.\par 
Starting from the FM side, the spin stiffness goes to zero in all directions as $\eta \rightarrow (\minus1)^{\minus}$, preserving the 4-fold symmetry of the ground state. However, similar to the stripe-type AFM order, the effective coupling between FM chains in the LO direction goes to zero, as shown in Fig.$~$\textcolor{blue}{3(b)}, and any TR component is steeply dispersive. Thus, the low-energy magnetic spectral weight is confined to a wall along the LO direction \big[Fig.$~$\textcolor{blue}{3(d)}\big], the scattering signature of a 1D system.\par  
 
 In summary, we show that CaCo$_{\mathrm{2}-y}$As$_{2}$ possesses spin fluctuations that are unique compared to \textit{A}Fe$_{2}$As$_{2}$ and SrCo$_{2}$As$_{2}$ in that it displays extreme spatial anisotropy. This extreme spatial anisotropy is due to the perfect magnetic frustration arising from the competing FM and AFM interactions and leads to effectively 1D behavior. Also, the value of the frustration parameter $\eta = \minus1.03(2)$ is in the region where the possibility of spin liquids are discussed\textcolor{blue}{\cite{Shannon_2004}}. Perfect magnetic frustration, spin liquids, and the exotic properties related to them are extensively discussed and realized in strongly-correlated (local moment) systems. However, the role of magnetic frustration in moderately/weakly correlated metals is poorly understood. In this regime, coupling of charge carriers to quantum spin fluctuations can lead to many interesting quantum phenomena, including unconventional superconductivity.  \cca\ is a very unusual metallic square-lattice compound in which nearly perfect frustration occurs. The challenge is to identify other potential candidates.  For example, one could dope insulators or semiconductors known to be near maximal frustration into the metallic state to generate materials that retain the frustration and in which novel electronic and/or magnetic phenomena may be found.\par

%\begin{table}[h]
%\caption{Fit parameter values and calculated values $\xi$ and $\xi_{\mathrm{LO}}$ from $\xi_{\mathrm{TR}}$ .}
%\centering
%\begin{tabular}{||p{1.7cm} | p{1.9cm} | p{1.7cm} | p{1.7cm} | p{1.7cm} ||}
%\hline \hline
%$\Gamma$ $(\mathrm {meV})$ & $\eta$ & $\xi$ $(\mathrm{\AA})$  & $\xi_{\mathrm{TR}}$ $(\mathrm{\AA})$ & $\xi_{\mathrm{LO}}$$(\mathrm{\AA})$ \\ [5pt]
%\hline
%22(5) & \minus1.010(8) & 3.37(3) & 4.79(4) & 0.47(10)\\ [5pt]
%\hline
%\end{tabular}
%\end{table}

%\begin{methods}

%\end{methods}

%\bibliography{CaCo2As2_v10}

\begin{acknowledgments}
We are grateful for assistance of G. S. Tucker (Paul Scherrer Institute), Abhishek Pandey (Texas A\&M University), Yongbin Lee (Ames Laboratory), Jong Keum (X-ray Laboratory, SNS, ORNL), Songxue Chi (TAX, HFIR, ORNL), and Rafael M. Fernandes (University of Minnesota).  Work at the Ames Laboratory was supported by the U. S. Department of Energy, Basic Energy Sciences, Division of Materials Sciences \& Engineering, under Contract No. DE-AC02-07CH11358. This research used resources at the Spallation Neutron Source, a DOE Office of Science User Facility operated by the Oak Ridge National Laboratory.
\end{acknowledgments}

\bibliography{CaCo2As2_arxiv}

%%%%%%%%%% Merge with supplemental materials %%%%%%%%%%
\pagebreak
\widetext
\begin{center}
	\textbf{\large Supplemental Material for Effective One-Dimensional Coupling in the Highly-Frustrated Square-Lattice Itinerant Magnet CaCo$_{\mathrm{2}-y}$As$_{2}$}
\end{center}
\setcounter{equation}{0}
\setcounter{figure}{0}
\renewcommand{\thefigure}{S\arabic{figure}}

\vspace{5mm}
\noindent\underline{\large\textbf{Heisenberg Model:}}\newline
The dispersion relation of the A-type AFM order, associated with the Heisenberg Hamiltonian consisting of nearest-neighbor ($\textit{J}_{1}$), next-nearest-neighbor ($\textit{J}_{2}$) and interlayer ($\textit{J}_{c}$) exchange parameters in the \textit{I}4/\textit{mmm} cell, is
\begin{equation}
\textit{E}(\textbf{q}) = \sqrt{A^2_{\mathrm{\textbf{q}}}\minus B^2_{\mathrm{\textbf{q}}}}\CommaPunct
\end{equation}
where \textbf{q} = \textbf{Q} \minus \textbf{Q}$_\mathrm{A}$ is the reduced wavevector and\newline 
\indent$\textit{A}_{\mathrm{\textbf{q}}} = \textit{S}\textit{J}_{c}\minus4\textit{S}(\textit{J}_{1}+\textit{J}_{2})+4\textit{S}\Big\{\dfrac{\textit{J}_{2}}{2}\big[\cos(2 \pi H)+\cos(2 \pi K)\big]+\textit{J}_{1}\cos(\pi H)\cos(\pi K)\Big\}$\newline
\indent$\textit{B}_{\mathrm{\textbf{q}}} = \textit{S}\textit{J}_{c}\cos(\pi L)$\newline
The neutron scattering cross-section can be written as \textcolor{blue}{\cite{Harriger_2011,Johnston_2010}}
\begin{equation}
\dfrac{\text{d}^2\sigma}{\text{d}\Omega \text{d}E}\ = A\dfrac{k_\mathrm{f}}{k_\mathrm{i}} \sum_{\alpha, \beta}(\delta_{\alpha\beta} \minus\hat{Q}_\alpha \hat{Q}_\beta)S^{\alpha \beta}.
\end{equation}
Here, \textit{A} is the overall scale factor, \textit{k$_\mathrm{i}$} and \textit{k$_\mathrm{f}$} are initial and final wavevectors, respectively, and $S^{\alpha \beta}$ is the response function or the dynamical structure factor describing spin-spin correlations. Considering only the transverse-component contributions to the neutron intensity, the response function can be described by the damped simple harmonic oscillator (DSHO) as\cite{Harriger_2011,Johnston_2010}
\begin{equation}
S(\textbf{Q}, E)\ = S_{\mathrm{eff}} \dfrac{(A_\mathrm{\textbf{q}} \minus B_\mathrm{\textbf{q}})}{1\minus e^{\minus E/k_{\mathrm{B}}T}}\dfrac{4}{\pi}\dfrac{\Gamma E}{\big[E^2\minus E^2(\textbf{q})\big]^2 + 4(\Gamma E)^2}\CommaPunct
\end{equation}
where \textit{k}$_{\mathrm{B}}$ is the Boltzmann constant, $\Gamma$ is the damping parameter, and \textit{S$_{\mathrm{eff}}$} is the effective spin.\par
\vspace{10mm}
\noindent\underline{\large\textbf{Estimation of $\textit{J}_{1}$ :}}\newline
From the dispersion relation of the A-type antiferromagnet, neglecting the interactions along \textit{c}-direction, $\textit{J}_{1}$ can be written in terms of the half width at half maximum $(\delta)$ along the TO direction at (\textit{H}, \textit{K}) $=$ (0.5, 0.5) as:
\begin{equation}
S|\textit{J}_{1}| = \dfrac{E}{4\pi^2\delta^2}\CommaPunct
\end{equation}
where $\delta$ is in r.l.u.$~$and from our data, for $\textit{E} = 55$ meV, we get $\delta = 0.11$ r.l.u. Substituting these values in equation (4), we get
\begin{equation}
S|\textit{J}_{1}| \approx 57.5 \mathrm{\ meV}.
\end{equation} 
\newpage
\noindent\underline{\large\textbf{Spin-wave spectrum calculated using the local-moment Heisenberg model with DSHO}}\newline
\begin{figure*}[htb]
	\vspace{-1em}
	\centering
	\hspace{-1.4em}
	\includegraphics[trim = 40mm 95mm 35mm 95mm,clip,width=17cm]{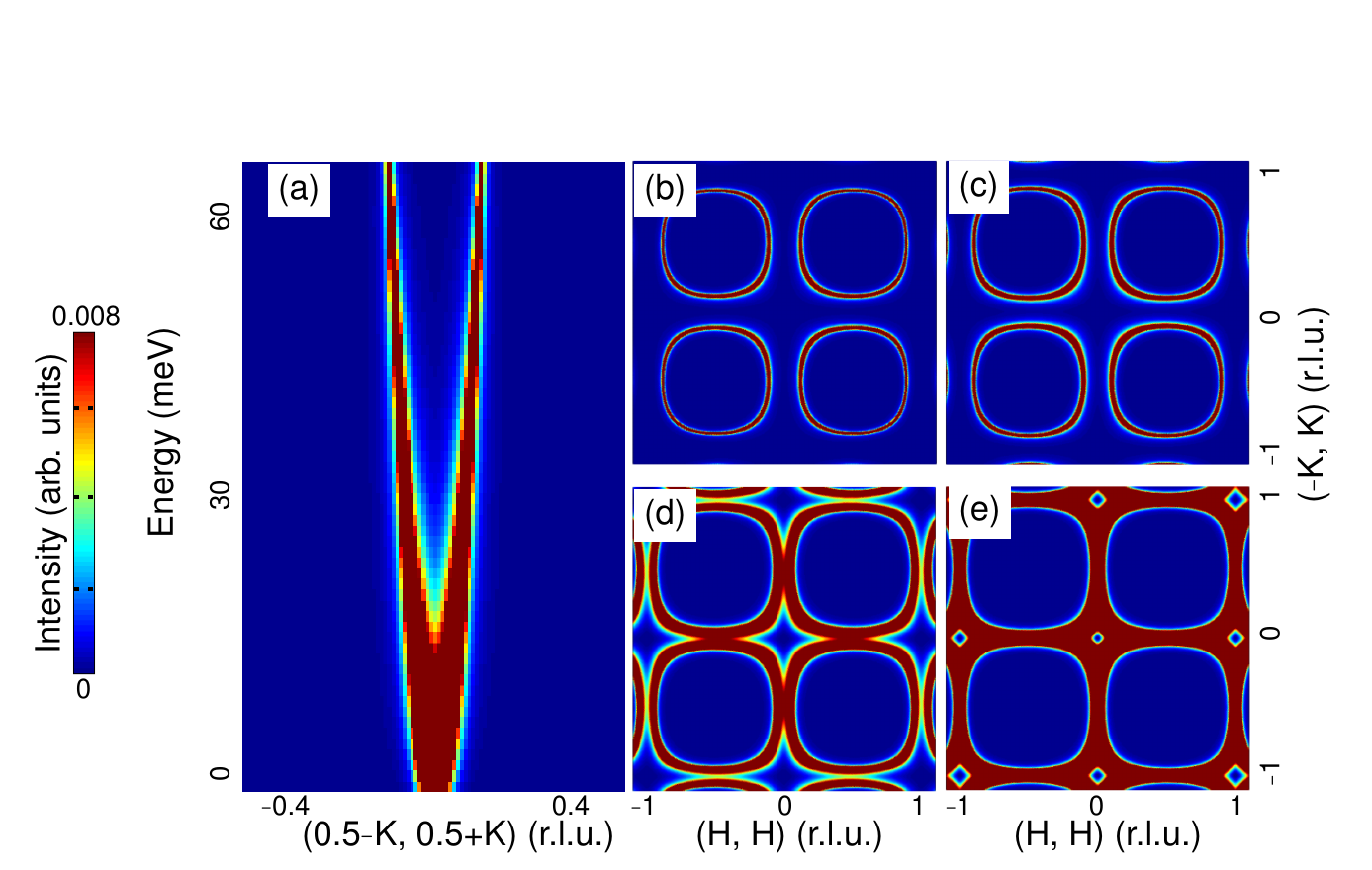}	
	%\captionsetup{labelformat=empty}
	\caption{\noindent(a) Slice showing the energy dependence of spin fluctuations along the TR direction. (b)--(e) Constant-energy slices for energy transfers of (b) 55 $\pm$ 5 meV, (c) 45 $\pm$ 5 meV, (d) 15 $\pm$ 5 meV, and (e) 5.5 $\pm$ 4.5 meV\@. Unlike the experimental data, the figures show a dispersion of the peak at the higher energy transfer with no intensity along the LO direction. The spectrum is obtained with values $S\textit{J}_{1} = \protect\minus50$ meV, $S\textit{J}_{2} = 24.4$ meV and $S\textit{J}_{c} = 0.0$ meV\@.}	
\end{figure*}
%\vspace{5mm}
\newpage
\noindent\underline{\large\textbf{Spatial anisotropy of the spin fluctuations for different magnetic ordering}}\newline
\begin{figure}[H]
	\begin{adjustwidth}{4mm}{}
		\vspace{0em}
		\centering

		\includegraphics[trim = 25mm 65mm 80mm 22mm,clip,width=10cm]{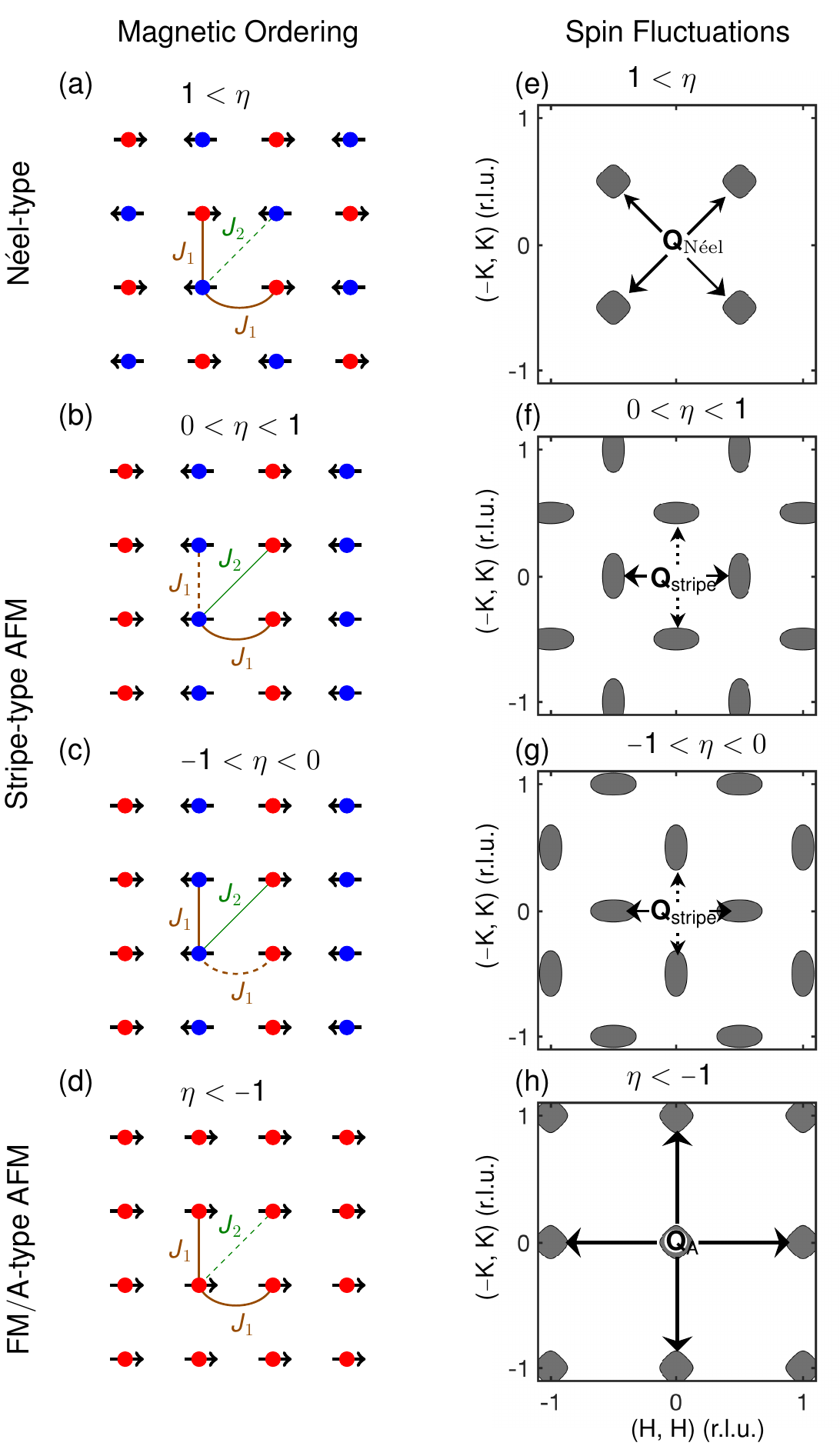}
		%\put(-180,83){{\color{black}\fontsize{9}{9}\fontfamily{phv}\selectfont\colorbox{white}{\ovalbox{$\textbf{Q}_{\mathrm{\text{A}}}$}}}}
		%\renewcommand{\sfdefault}{phv}
		%\renewcommand{\familydefault}{\sfdefault}\caption{\noindent\fontsize{11pt}{0pt}\selectfont{
		\caption{\noindent{(a)--(d) Real space arrangement of spins on a square lattice for: (a) N$\acute{\mathrm{\text{e}}}$el-type, (b),(c) stripe-type AFM and (d) FM/A-type order. Red and blue circles represent transition metals on different magnetic sublattices. Magnetic NN (\textit{J}$_{1}$) and NNN (${\mathrm{\textit{J}}}_{2}$) interactions are shown where \textit{J}$_{1}$ is AFM in (a) and (b) and FM in (c) and (d) and \textit{J}$_{2}$ is AFM. Dashed lines indicate frustrated interactions when \textit{J}$_{2}$ is AFM. (e)--(h), Spread and positions of the spin fluctuations in reciprocal space for finite neutron energy transfer for the corresponding magnetic order. Dotted arrows in (f) and (g) point to the spin fluctuations of the domains which have magnetic ordering similar to (b) and (c), respectively, but with alternating chains of blue and red circles along the perpendicular direction. N$\acute{\mathrm{\text{e}}}$el and FM/A-type fluctuations have four-fold anisotropy. The spin fluctuations corresponding to the different magnetic orderings show differences in spatial anisotropy.}}
	\end{adjustwidth}
\end{figure}

\newpage
\noindent\underline{\large\textbf{Slices of raw data (Inelastic Neutron Scattering)}}\newline
\vspace{-5em}
\begin{figure}[htb]
	\includegraphics[trim = 50mm 90mm 50mm 60mm,clip,width=16cm]{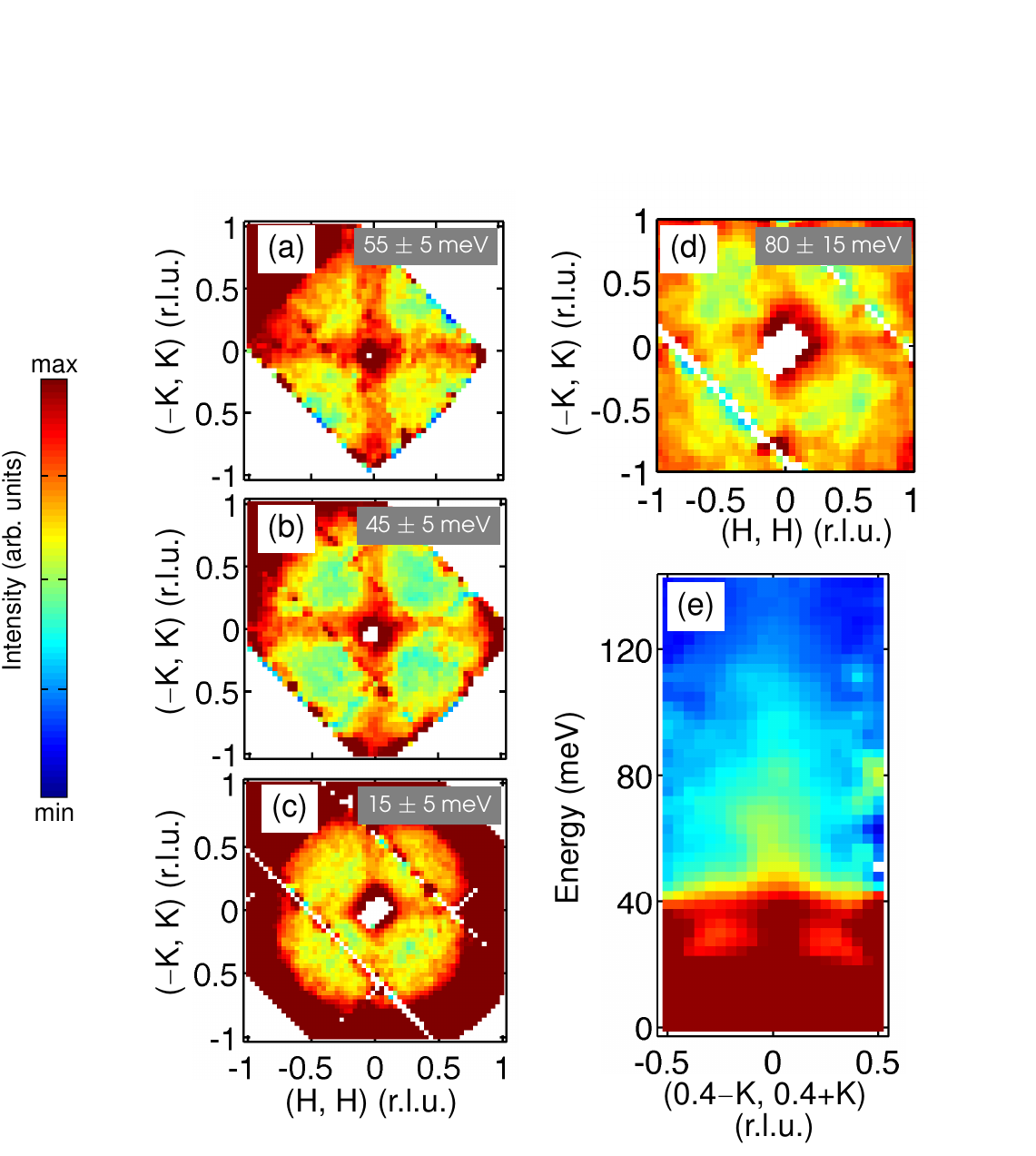}
	%\captionsetup{labelformat=empty}
	\caption{\noindent{Raw data measured on CaCo$_{2-y}$As$_{2}$ at \textit{T} $ = 8$ K at ARCS using INS}. (a)--(c) Data measured with \textit {E$_{\mathrm{i}}$} $= 75$ meV and (d) and (e) with  \textit{E$_{\mathrm{i}}$} $ = 250$ meV. (a)--(c), Constant-energy slices of the raw data in the (H, H)-(\protect\minus K, K) plane averaged over energy ranges of (a) 50--60 meV, (b) 40--50 meV and (c) 10--20 meV. (d) Constant-energy slice of the data in the (H, H)-(\protect\minus K, K) plane averaged over an energy range of 65--95 meV. (e) Transverse slice of the data along the $[$\protect\minus K, K$]$ direction through  (0.4, 0.4) showing the presence of the spin fluctuations up to 120 meV.}
\end{figure}

\newpage
\noindent\underline{\large\textbf{Evolution of the spin fluctuations with the frustration parameter ($\eta$)}}
\begin{figure}[htb]
	\begin{adjustwidth}{2mm}{}
		%\begin{tabular}{@{}l@{}l@{}}
		\centering	
		\includegraphics[trim = 10mm 52mm 25mm 30mm,clip,width=12cm]{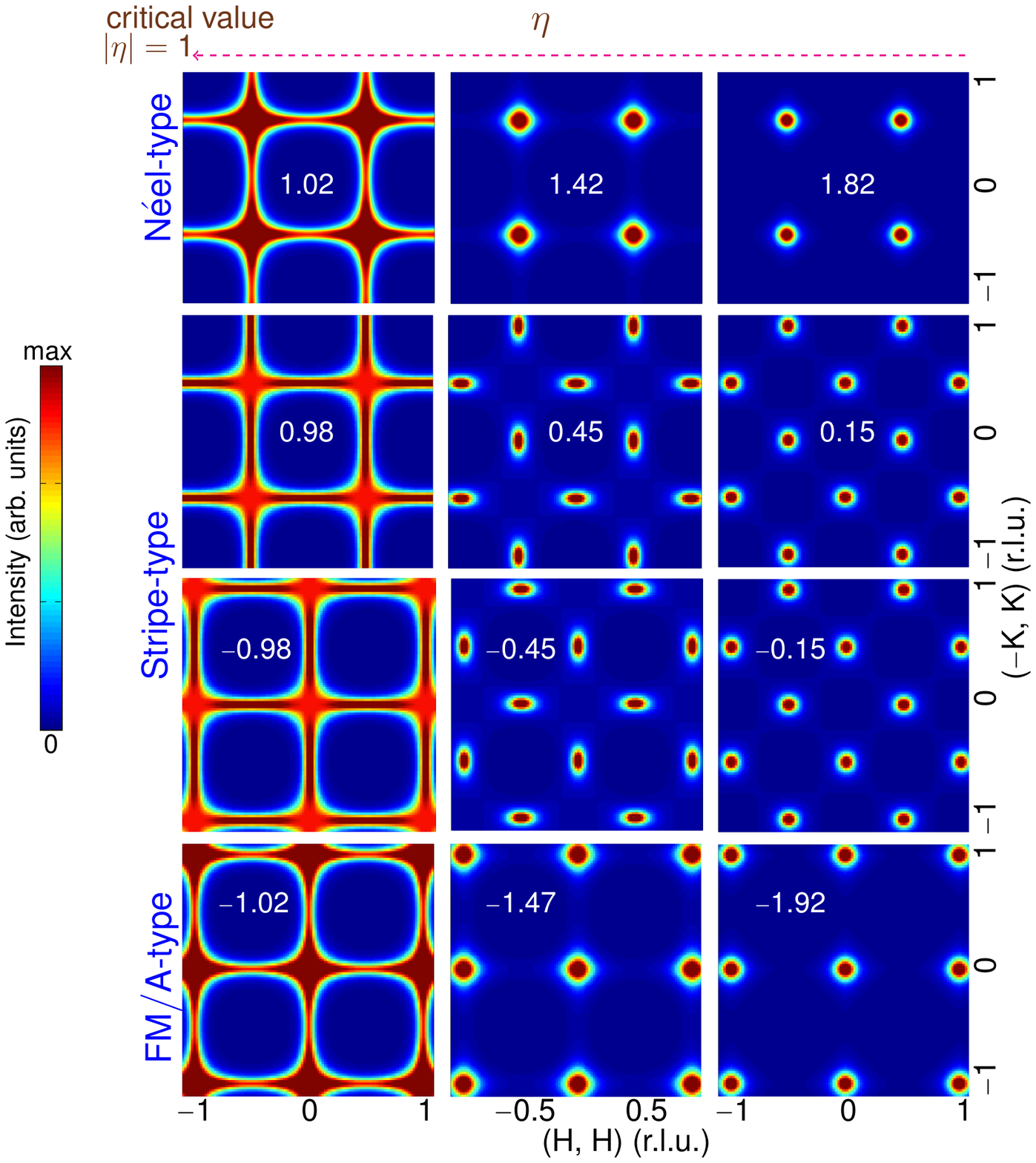}\vspace{0em}
		%	\vspace{0em}\includegraphics[trim = 0mm 0mm 0mm 0mm,clip,width=20cm]{fig3_aftrrob}
		%\renewcommand{\sfdefault}{phv}
		%\renewcommand{\familydefault}{\sfdefault}\caption{\noindent\fontsize{11pt}{0pt}\selectfont{
		\caption{\noindent{Constant energy slices showing INS cross-section calculated using the diffusive model for different magnetic ordering. The spin fluctuations become extremely anisotropic for all ordering with the frustration parameter $|\eta| \rightarrow 1$ (maximum frustration). This supports the fact that extremely anisotropic spin fluctuations are consequence of the system being in a highly-frustrated state. For maximum frustration $\eta =\protect\minus 1$ ($1$), the spin fluctuations for stripe-type and FM/A-type (N$\acute{\mathrm{\text{e}}}$el-type and stripe-type) become identical. The constant-energy slices are averaged over the energy ranges of 10--20 meV\@.}}
	\end{adjustwidth}
\end{figure}  
\clearpage
%\bibliography{supplement}

\end{document}